\documentclass[aps,10pt,prd,citeautoscript,longbibliography,superscriptaddress,twocolumn,nofootinbib,floatfix]{revtex4-1}
\usepackage[T1]{fontenc}
\usepackage{color}
\usepackage{graphicx}
\usepackage{amsmath,amssymb,amsfonts,bm,dsfont,units}
\usepackage{float}

\usepackage{microtype}
\usepackage{multirow}
\usepackage[dvipsnames]{xcolor}

\usepackage[caption=false]{subfig}

\usepackage[colorlinks=true,citecolor=blue,linkcolor=blue,urlcolor=blue]{hyperref}
\usepackage[capitalize,nameinlink]{cleveref}

\newcommand{\redmond}{Microsoft Quantum, Redmond, Washington 98052, USA}
\newcommand{\santabarbara}{Microsoft Station Q, Santa Barbara, California 93106 USA}

\begin{document}

\title{Performance of planar Floquet codes with Majorana-based qubits}

\author{Adam Paetznick}
\affiliation{\redmond}
\author{Christina Knapp}
\affiliation{\santabarbara}
\author{Nicolas Delfosse}
\affiliation{\redmond}
\author{Bela Bauer}
\affiliation{\santabarbara}
\author{Jeongwan Haah}
\affiliation{\redmond}
\author{Matthew B. Hastings}
\affiliation{\santabarbara}
\affiliation{\redmond}
\author{Marcus P. da Silva}
\affiliation{\redmond}

\begin{abstract}
Quantum error correction is crucial for any quantum computing platform to achieve truly scalable quantum computation. The surface code and its variants have been considered the most promising quantum error correction scheme due to their high threshold, low overhead, and relatively simple structure that can naturally be implemented in many existing qubit architectures, such as superconducting qubits.
The recent development of Floquet codes by Hastings and Haah offers another promising approach. By going beyond the usual paradigm of stabilizer codes, Floquet codes achieve similar performance while being constructed entirely from two-qubit measurements. This makes them particularly suitable for platforms where two-qubit measurements can be implemented directly, such as the measurement-only topological qubits based on Majorana zero modes (MZMs) proposed by Karzig et al. Here, we explain how two variants of Floquet codes can be implemented on MZM-based architectures without any auxiliary qubits for syndrome measurement and with shallow syndrome extraction sequences. 
We then numerically demonstrate their favorable performance. In particular, we show that they improve the threshold for scalable quantum computation in MZM-based systems by an order of magnitude, and significantly reduce space and time overheads below threshold.
\end{abstract}

\maketitle
\date{}

% \tableofcontents

\section{Introduction}
For over a decade the surface code~\cite{Kitaev1997,Bravyi1998,Freedman2001,dennis2002} has been the de-facto standard for scalable fault-tolerant quantum computing architectures. Its high threshold and local support on a two-dimensional planar lattice make it highly appealing for a wide variety of plausible physical implementations~\cite{Fowler2012,Bermudez2017,Cai2019,Chamberland2020,Bombin2021}. However, despite promising progress~\cite{Kelly15,Takita16,Harper19,Anderson20,Chen21,Marques21,Erhard21,Gong21}, full demonstration of a logical qubit encoded in a surface code patch has thus far remained elusive even for leading experimental platforms.

One qubit platform that promises to reach error rates far below what is required for the surface code is topological quantum computation. Topological qubits are endowed with topological protection, which can be thought of as hardware-level error protection that is able to suppress the effect of any local error exponentially in physical parameters such as the size of the system and the spectral gap. While many different physical systems exhibit topological order and could in principle be used for topological quantum computation~\cite{nayak2008nonabelian}, one of the most promising implementations are Majorana zero modes (MZMs)~\cite{Read00,Kitaev01} in semiconductor-superconductor heterostructures~\cite{Alicea12,Lutchyn18}. A particularly promising scheme to operate such qubits is measurement-only topological quantum computation~\cite{Bonderson08}, which avoids having to physically move topological excitations and instead performs all manipulations via (in the case of MZMs) joint fermion parity measurements on small numbers of MZMs. Specific proposals for such qubits were put forward in Ref.~\onlinecite{Karzig2017}. The natural operations of these qubits are measurements of multi-qubit Pauli operators in a collection of adjacent qubits. Exactly which Pauli operators are available depends on the details of the physical layout, which can thus be tailored to a particular application.

As stabilizer codes are constructed from multi-qubit Pauli measurements, it would seem that these MZM-based measurement-only topological qubits are ideally suited for the implementation of the surface code. Indeed, several possible approaches have been discussed in the literature, including direct measurement of the weight-4 Pauli operators~\cite{Plugge16}, generalizations thereof~\cite{Litinski18}, as well as circuits that implement the stabilizers as a sequence of weight-two measurements~\cite{Chao2020,Tran20}. However, these implementations either rely on operations that are challenging to implement physically or suffer from overhead in time and space that is much larger than implementations of the surface code based on, say, superconducting qubits.

Recently, this situation was remedied by a new class of codes that also have local support on a two-dimensional lattice~\cite{Hastings2021,Haah2021}.
These so-called ``Floquet codes'' can act as a fault-tolerant quantum memory by way of a time-ordered sequence of two-qubit Pauli measurements. Compared to the surface code, which is fundamentally constructed from four-qubit Pauli measurements, this obviates the need for compiling four-qubit Pauli measurements into either a sequence of two-qubit Clifford gates and single-qubit measurements, or one- and two-qubit Pauli measurements. The codes can be formulated on any face-three-colorable lattice and thus afford significant flexibility in the physical layout.
A natural choice of lattice is the honeycomb lattice. The honeycomb Floquet code on a torus was shown in Ref.~\onlinecite{Gidney2021} to have highly competitive thresholds and logical error rates. However, due to the boundary conditions, it is impractical for large-scale implementation on a physical plane. This was remedied in a follow-up~\cite{Haah2021}, where the authors propose a simple set of planar boundary conditions for the honeycomb code.

Here, we reconsider this planar honeycomb code and additionally introduce a Floquet code on the 4.8.8 lattice (also known as truncated-square or square-octagon lattice). For both the 4.8.8 code and the honeycomb code, we show that the bulk operations can be naturally implemented in an array of so-called \textit{tetrons}~\cite{plugge2017majorana,Karzig2017}, a particular variant of measurement-only MZM-based topological qubits.  This implementation uses only the most natural set of short-ranged measurements for this qubit platform. We then introduce boundary conditions for both the honeycomb and the 4.8.8 code that can be implemented directly in these tetron arrays, and which lead to physical implementations that require no auxiliary qubits and use only one physical operation per check operator.

Crucially, through numerical simulations of these planar Floquet codes, we demonstrate that they significantly outperform the surface code for the topological platform (or any other platform where two-qubit Pauli measurements are directly available). We find that the threshold improves from around 0.2\%~\cite{Chao2020} to above 1\% (see~\cref{table:thresholds}). 
Furthermore, we find substantial time and space savings depending on physical and logical target error rates. For example, Floquet codes offer a time savings between $5$--$10\times$ and a space savings up to $5\times$ for target logical error rate $10^{-12}$ and physical error rates between $10^{-6}$ and $10^{-3}$. These improvements are largely due to the comparatively high thresholds and more time-efficient syndrome measurements.

\section{Review of Floquet codes on the honeycomb lattice}
\label{sec:honeycomb-codes}

\begin{figure*}
\hfill
\subfloat[honeycomb unit]{\label{fig:honeycomb-unit}
  \includegraphics[width=.15\textwidth]{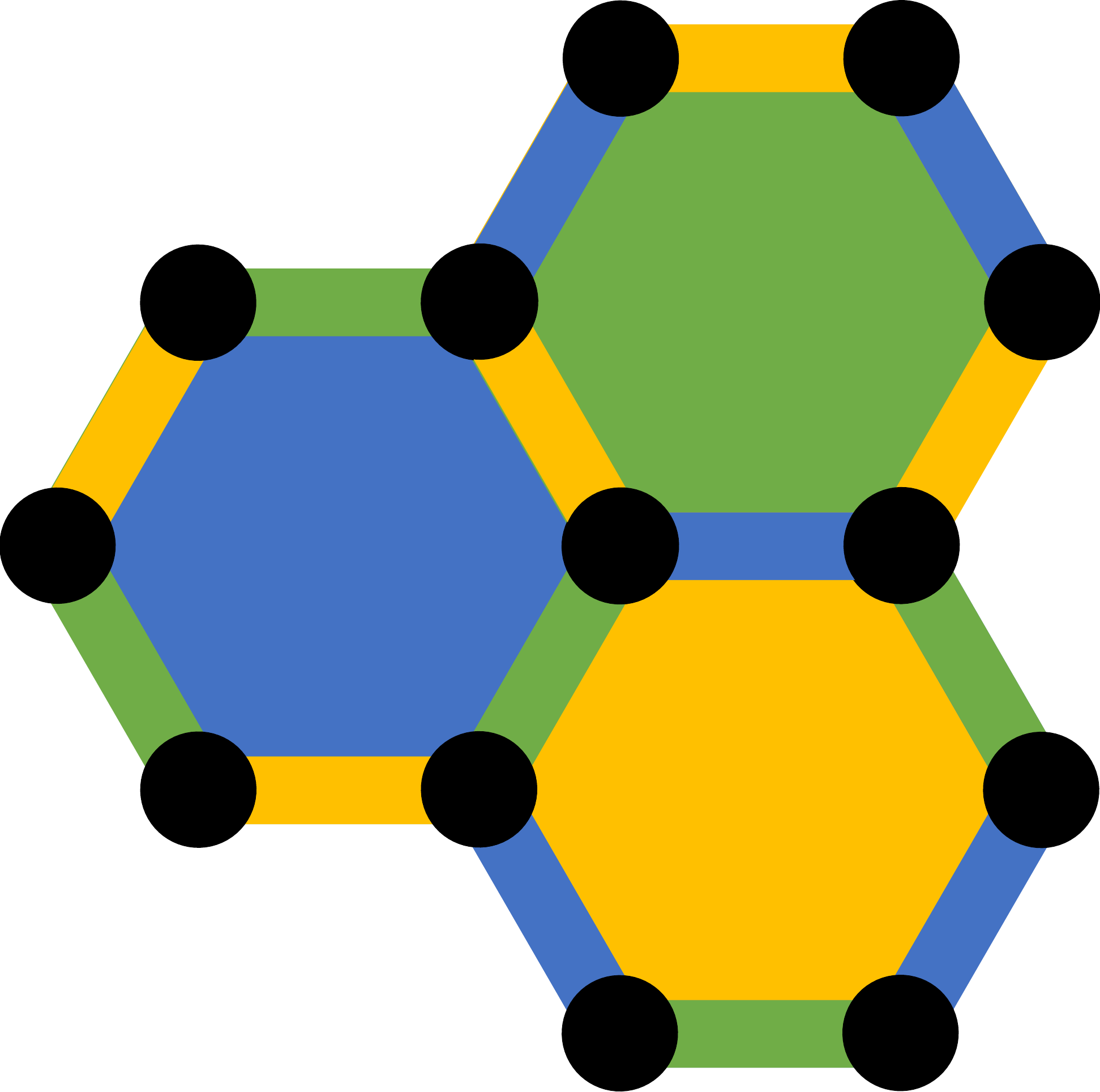}
}
\hfill
\subfloat[parallelogram]{
  \includegraphics[width=.5\textwidth]{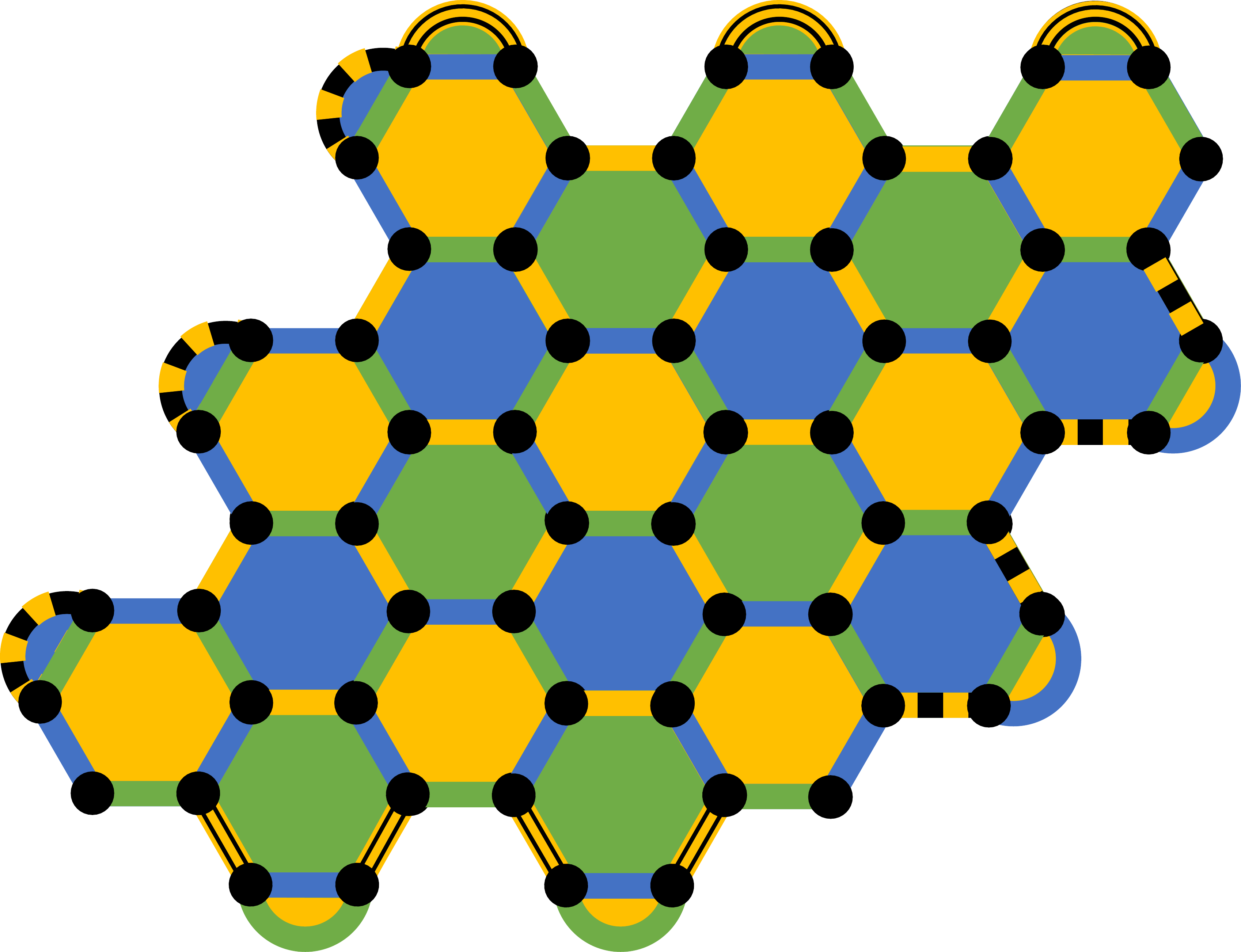}
  \label{fig:honeycomb-parallelogram}
}
\hfill
\caption{ 
(a) Tiling unit for the honeycomb code.  Qubits are indicated by black dots.
We assign a check operator $ZZ$ for each horizontal edge,
$XX$ for each edge with positive slope, 
and $YY$ for each edge with negative slope.
(b) Parallelogram boundary conditions for the honeycomb code. Solid color edges are measured in a period-six sequence (yellow, blue, green, yellow, green, blue). Curved edges are Pauli operators chosen such that the triple of check operators meeting at a vertex are pairwise anticommuting. (The degree-four vertex in the top left corner is treated specially, see~\cref{tab:honeycomg-isg}.)  For example, the curved green edges along the bottom are assigned $XY$ operators (from left to right). The striped black-yellow edges, bordering green faces, are measured during the first yellow round, while the dashed black-yellow edges, bordering blue faces, are measured during the second yellow round.
}
\end{figure*}

The honeycomb code, originally introduced in Ref.~\cite{Hastings2021}, is defined by a time-ordered sequence of weight-two Pauli operators supported on vertices of a hexagonal lattice. Each operator, or ``check,'' corresponds to an edge in the lattice. A family of honeycomb codes can be constructed by tiling some number of ``units'', which is a set of three adjacent faces as shown in~\cref{fig:honeycomb-unit}. The operators are assigned such that all operators supported at a given vertex anticommute.
A specific choice for the honeycomb lattice that satisfies these conditions is to assign $ZZ$ to horizontal edges, $XX$ to edges with positive slope, and $YY$ to edges with negative slope.
To define the order in which operators are measured, it is furthermore required to assign a color \{blue, yellow, green\} to each edge such that no two edges touching the same vertex have the same color.  Then, each face is surrounded by two colors of edges, and we choose to color each face by the lone remaining color; see~\cref{fig:honeycomb-unit} for an illustration.
For a lattice with periodic boundary conditions (a torus), a tiling using this unit is all that is required.  The check measurements are then time-ordered by color: first yellow, then blue, then green.  This period-three sequence forms one \emph{round} of syndrome extraction for the honeycomb code.  A number of rounds proportional to the distance is required for a fault-tolerant quantum memory.

Other boundary conditions require modifications, as discussed in detail in Ref.~\cite{Haah2021}.  One set of boundary conditions forms a parallelogram shape as shown in~\cref{fig:honeycomb-parallelogram}.  First, a set of units is tiled to form the bulk of the parallelogram.  Then, additional qubits and edges are added around the perimeter.  Some of the faces around the boundary are incomplete, resulting in 2-gons instead of hexagons.\footnote{The parallelogram patch shown in Ref.~\cite{Haah2021} used slightly different boundary conditions that also included 4-gons along the boundaries.
This original patch has almost identical performance to the one shown in \cref{fig:honeycomb-parallelogram}, but involves long-distance measurements when mapped to a square qubit lattice and is thus disfavored in our physical layout.}
The parallelogram boundaries also require a modification to the time-ordering of the check measurements.  Instead of measuring checks in the period-three sequence (yellow, blue, green), the edges are measured in a period-six sequence (yellow, blue, green, yellow, green, blue). Most of the edges, including all of the blue and green edges, are measured twice during this sequence.  The black-yellow dashed edges in~\cref{fig:honeycomb-parallelogram} are measured only once.

\subsection{Decoding graph construction and effective code distances}

The sequence of check measurements defines a corresponding sequence of stabilizer groups.  The instantaneous stabilizer group at a given time step is the group of Pauli operators obtained by projecting onto each check operator up to and including the chosen time step.  This includes all of the check measurements at that time step, and after at least four time steps, the Pauli operators supported on each of the hexagonal faces.  The instantaneous logical operators are the normalizers of the instantaneous stabilizer group, modulo the instantaneous stabilizer group.

On a torus (i.e., without boundaries), the instantaneous stabilizer group consists of the plaquette stabilizers around the hexagons and the check operators at the moment. With boundaries, the instantaneous stabilizer group possesses new types of elements (transient stabilizers), in addition to perpetual stabilizers associated with the hexagons, 4-gons, and 2-gons. First, when we measure yellow checks after green checks, we do not measure black-yellow striped checks bordering green faces. Hence, the green checks at the bottom of~\cref{fig:honeycomb-parallelogram} remain in the instantaneous stabilizer group. In the next step, we measure the green checks again and, in the absence of errors, the outcomes of the green checks at the bottom boundary are deterministic. This gives a node in the decoding graph, inferred by 2 green check outcomes. It is contrasted to a decoding graph node corresponding to a hexagon, which is inferred by 12 check outcomes. Second, similarly, the instantaneous stabilizer group at the yellow step (that follows a green step) contains the product of the outcomes of the two green checks at the top of~\cref{fig:honeycomb-parallelogram} which are separated by one yellow edge. Since we measure the green checks at the next step, we obtain a decoding graph node that is inferred by 4 green check outcomes. Finally, the top-left corner is exceptional, where two 2-gons do not carry perpetual stabilizers. The top-left qubit and its two neighbors support a subgroup of the instantaneous stabilizer group (ISG). This subgroup has period six up to signs (see \cref{tab:honeycomg-isg} for concrete details).
In two of the six rounds, the ISG subgroup remains unchanged and the corresponding check outcome is deterministic (in the absence of errors).
Whenever there is a deterministic outcome, there is a corresponding decoding graph node.
\begin{table*}
\centering
\begin{tabular}{c|c|c|c}
    \hline
    Round & Check operator & In steady state, ISG contains up to signs & Outcome deterministic?\\
    \hline
    Y & $IYX$ & $IZZ$, $IYX$ & \\
    B & $IZZ$ & $IZZ$, $IYX$ & Yes\\
    G & $XXI$ & $XXI$, $IXY$ & \\
    Y & $ZYI$ & $XXI$, $ZYI$ & \\
    G & $XXI$ & $XXI$, $ZYI$ & Yes\\
    B & $IZZ$ & $YZI$, $IZZ$ & \\
    \hline
\end{tabular}
\caption{
\label{tab:honeycomg-isg} 
Elements of the instantaneous stabilizer group, restricted to the three qubits in the top-left of~\cref{fig:honeycomb-parallelogram}.  In two of the six rounds the restricted ISG remains unchanged and the check outcome for that round is deterministic (in the absence of errors).  For operators in the table, the three qubits are ordered by starting with the leftmost of the three, and moving clockwise.
}
\end{table*}

When tiled in the fashion shown in~\cref{fig:honeycomb-parallelogram}, either on the torus or the parallelogram, the minimum weight of the logical operators is given by $2\ell$ where $\ell^2$ is the number of units in the tiling.  That is, the minimum weight corresponds to the length of the linear dimension, as measured in the units of~\cref{fig:honeycomb-unit}.

The effective distance of the code depends on the circuit and corresponding noise model used to implement the check measurements. If errors occur independently on each qubit, the distance is equal to the minimum weight $2\ell$ of the logical operators.  Under other noise models, the effective distance can be lower and is defined to be the minimum number of errors in that noise model (where an error may act on several qubits) which may cause an undetectable logical error.  For the noise model we consider in this paper, where errors may occur on pairs of qubits, the effective distance is cut in half to $\ell$, see~\cref{sec:noise-model}.

Note that in Floquet codes, it is possible for measurement errors alone (i.e., a bit flip of the measurement outcome without any error on the qubits of the code) to lead to an undetectable logical error, and with respect to such errors the distance in the parallelogram is equal to $2\ell-1$.  In \cref{fig:honeycomb-parallelogram}, there is a set of $5$ measurement errors, all in a line starting at a solid yellow edge on the left boundary and ending at a dashed yellow edge on the right boundary (four edges of which are parallel to each other), which leads to an undetectable logical error. 

\subsection{Logical operator dynamics}\label{sec:logical-operator-dynamics}
A distinctive feature of Floquet codes is that both the stabilizer group \emph{and} the logical operators change over time.
Check measurements for a given time step may anticommute with representatives of the instantaneous logical operators from the previous time step.  The logical operators can always be multiplied by stabilizers from the previous instantaneous stabilizer group in order to resolve this issue.  This changes both the support and relative phase of the logical operators.  

For example, let $X_t$ be a logical operator immediately following time $t$.
Let $S_t$ be a check operator measured at time $t$, and $\phi \in \{\pm 1\}$ be a measurement-dependent phase such that $\phi S_t$ is in the instantaneous stabilizer group immediately following time $t$.  Let $S_{t+1}$ be a check operator measured at time $t+1$ such that $\{X_t, S_{t+1}\} = \{S_t, S_{t+1}\} = 0$.
We may then choose $X_{t+1} = \phi S_t X_t$ so that $[X_{t+1}, S_{t+1}] = 0$.  Note that $X_{t+1}$ picks up a phase $\phi$ relative to $X_t$.  The phase $\phi$ is dependent on the measurement outcome $\{\pm 1\}$ of the check measurement for $S_t$.

At any moment in time, each logical operator is described by a Pauli operator on which it is supported, and a $\pm 1$ phase.  This phase is the product of a subset of outcomes from the history of check measurements.  We call this set of measurements the \emph{logical phase conditions}, which are crucial when computing the logical effect of an error (see~\cref{sec:calculating-logical-errors}).

\section{Floquet codes on the 4.8.8 (square-octagon) lattice}\label{sec:488-codes}

\begin{figure*}
\hfill
\subfloat[unit tile]{
    \includegraphics[width=.2\linewidth]{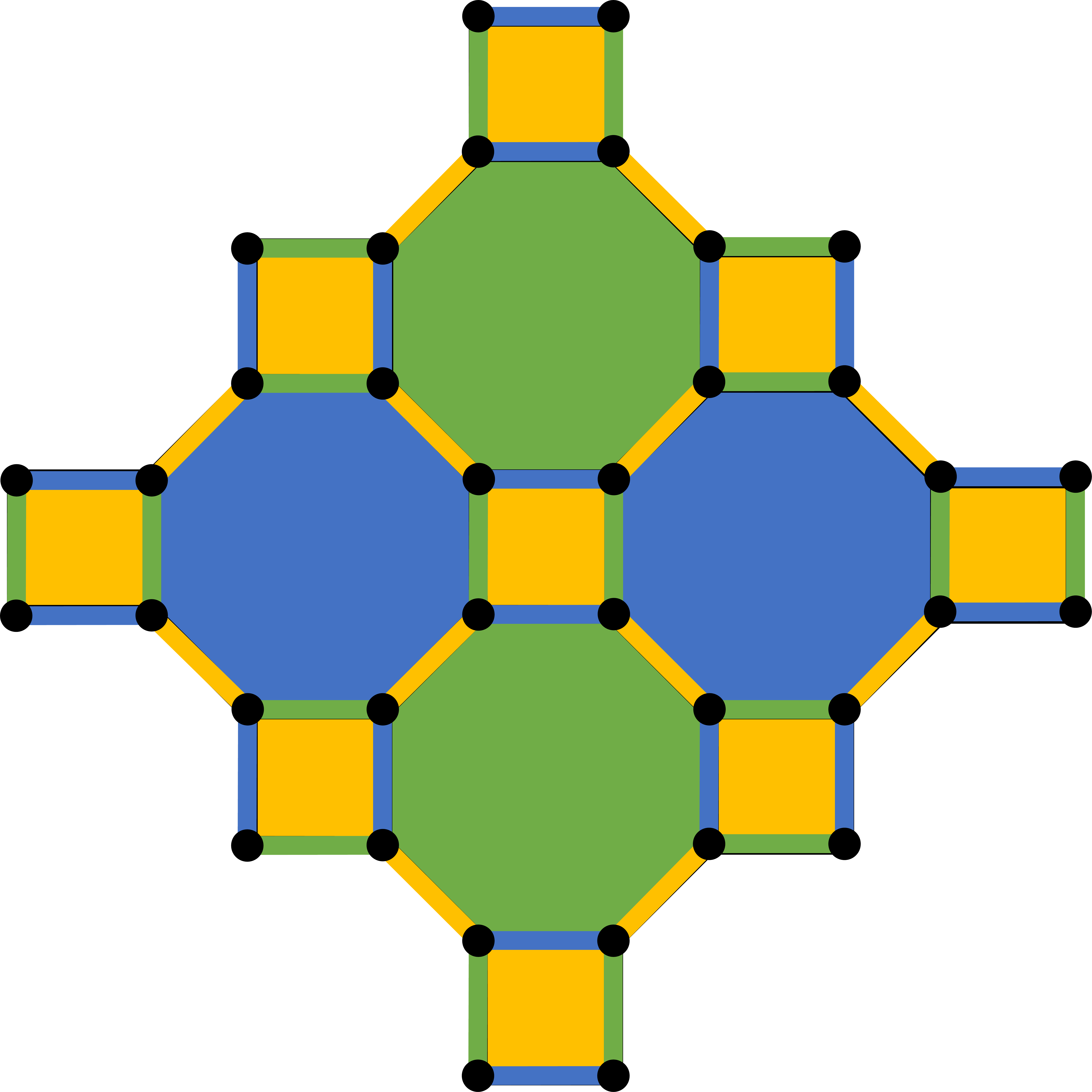}
    \label{fig: 488-unit-tile}
}
\hfill
\subfloat[rectangle]{
    \includegraphics[width=.45\linewidth]{488-parallelogram-2x2.pdf}
    \label{fig:488-rectangle-2x2}
}
\caption{
(a) Tiling unit for the 4.8.8 Floquet code. Like~\cref{fig:honeycomb-unit}, qubits are indicated by black dots.
Every diagonal edge is $YY$, horizontal $ZZ$, and vertical $XX$.
(b) Rectangular boundary conditions for the 4.8.8 Floquet code. Like the honeycomb code, solid color edges are measured in a period-six sequence (yellow, blue, green, yellow, green, blue). Dashed and striped black-yellow edges are the same Pauli type as solid yellow edges, but are measured at different rounds in the measurement sequence.  Dashed edges, incident to green faces, are measured during the first yellow round. Striped edges, incident to blue faces, are measured during the second yellow round.
}
\end{figure*}

The honeycomb code is a promising candidate for fault-tolerant quantum error correction because of its logical error performance and its layout on a two-dimensional planar grid.  One drawback, however, is that the effective distance of the code is cut in half by correlations of the errors on two-body check measurements, as in the noise model we consider here.  Fortunately, this limitation can be avoided by using the Floquet code model on the 4.8.8 (truncated square or square-octagon) lattice.

Like the honeycomb lattice, the 4.8.8 lattice is three-colorable and can therefore be used to construct a similar family of Floquet codes. The basic unit is shown in~\cref{fig: 488-unit-tile}, which can be used to tile the torus. Qubits are again placed on the vertices of the lattice, each of which is incident to three edges of different colors \{blue, yellow, green\}. 
We assign $XX$ to each vertical edge (colored green or blue), $YY$ to each diagonal edge (colored yellow), and $ZZ$ to each horizontal edge (colored green or blue).
(This choice makes Pauli frames translation invariant in the underlying physical qubits.) Like-colored edges are measured simultaneously and colors are alternated in a time-ordered sequence.
Note that the types of decoding graph nodes for the 4.8.8 Floquet code are conceptually simpler than those of honeycomb Floquet code. There are perpetual stabilizers associated with each plaquette (8-, 4-, and 2-gons), leading to decoding graph nodes inferred by 16, 8, or 4 check outcomes. In addition, there are decoding graph nodes inferred by just 2 check outcomes along the boundary.

To define the code with open boundaries, again a choice of boundary conditions is required. A convenient choice is shown in~\cref{fig:488-rectangle-2x2}.
Time-ordering of the check measurements is also analogous to the honeycomb code.  On the torus the period-three sequence (yellow, blue, green) is sufficient.  On the plane, we use the period-six sequence (yellow, blue, green, yellow, green, blue).  Black-yellow dashed edges are each measured only once during the period-six sequence.
The planar check sequence prescribed above can be modified by permutations of the colors.  For example (blue, yellow, green, blue, green, yellow).  Because of the asymmetry of the faces in the 4.8.8 lattice, the choice of colors is not entirely arbitrary.  It is possible that other permutations may change logical performance.

If errors occur independently on each qubit, then, like the honeycomb, the distance is equal to the minimum weight~$d$ of the logical operators.
Unlike the honeycomb, the effective distance of the 4.8.8 code remains unchanged if the noise model includes two-qubit errors on the check measurements.  This property leads to better spatial overhead, as discussed in~\cref{sec:low-error-rate-performance}, under a two-qubit correlated noise model. 
The essential reason for this is that every two-qubit operator on the support of a check operator intersects any minimum-weight logical operator only in one qubit. 

One can formally show that the effective distance is still $d$ under two-qubit errors on the check measurements by the following consideration.
As shown in Ref.~\onlinecite{Hastings2021}, the decoding graph can be constructed under a simplified error model in which a single-qubit Pauli error~$P$ may be introduced only if it immediately follows the measurement of a check operator~$P\otimes P$. Any Pauli error configuration in spacetime defines a $1$-chain in this decoding graph, and a logical operator corresponds to a path from one boundary to the opposite boundary of the decoding graph. Then, it suffices to show that no two-qubit error on the support of any check operator can advance the end point of the $1$-chain by more than graph distance one from a boundary plane of the decoding graph.

\section{Physical layout}
\label{sec:physical-layout}

\begin{figure*}[t]
\includegraphics[width=1.99\columnwidth]{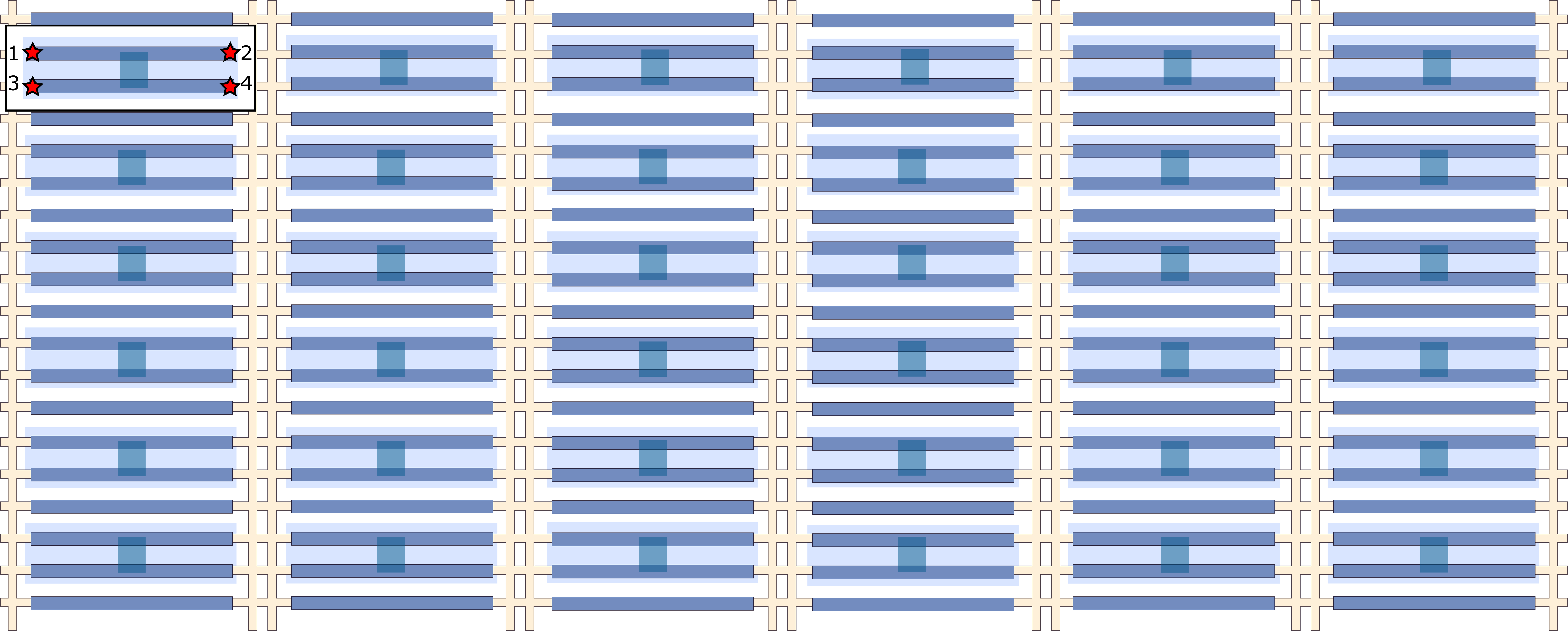}
\caption{\label{fig:phys-layout} 
Square lattice of tetrons used to implement the honeycomb and 4.8.8 Floquet codes.
Topological superconducting wires (dark blue) have a MZM at either end.
Qubit islands (light-blue shaded) correspond to two parallel topological wires joined by a trivial superconducting backbone (teal), with MZMs  (red stars) labeled according to the box in the upper left.
Rows of tetrons are separated by coherent links (floating topological wires).
Neighboring qubit islands are connected by semiconducting segments (tan), with two semiconducting columns separating each column of qubits.
}
\end{figure*}

For the physical implementation of these codes, we consider a square grid of so-called \textit{tetron} qubits. We briefly review the architecture here, see Ref.~\cite{Karzig2017} for a more detailed description. The tetron has two parallel $p$-wave superconducting wires~\cite{Kitaev01,Motrunich01} connected by a trivial superconducting wire in the center. Together, these form a qubit island with a finite charging energy, which fixes the total fermion number in the ground state and suppresses quasi-particle poisoning events.
Each topological wire has a Majorana zero mode (MZM) at either end, described by self-adjoint fermion operators $\gamma_j$, $j\in\{1,2,3,4\}$ labeled according to the box in the upper left of~\cref{fig:phys-layout}. These satisfy the canonical anticommutation relations $\lbrace \gamma_i, \gamma_j \rbrace = 2\delta_{ij}$. Fixing the total parity to be even, $-\gamma_1\gamma_2\gamma_3\gamma_4 =1$, these four Majorana zero modes form a two-dimensional degenerate ground state that can be used to encode a qubit. The single-Pauli operators are given by Majorana bilinears as
\begin{align}\label{eq:basis}
    i\gamma_1\gamma_2 &= i\gamma_3 \gamma_4 = X,
    \\ i\gamma_1\gamma_4 &= i\gamma_2\gamma_3 = Y,
    \\ i\gamma_1\gamma_3 &= -i\gamma_2\gamma_4 = Z.
\end{align}
The fact that each Pauli operator has two equivalent representations as Majorana bilinear will turn out to be crucial for an efficient implementation of Floquet codes. Two-qubit Pauli operators can be inferred straightforwardly and each corresponds to four equivalent products of four Majorana operators.

\begin{figure*}[t]
    \centering
    \includegraphics[width=1.8\columnwidth]{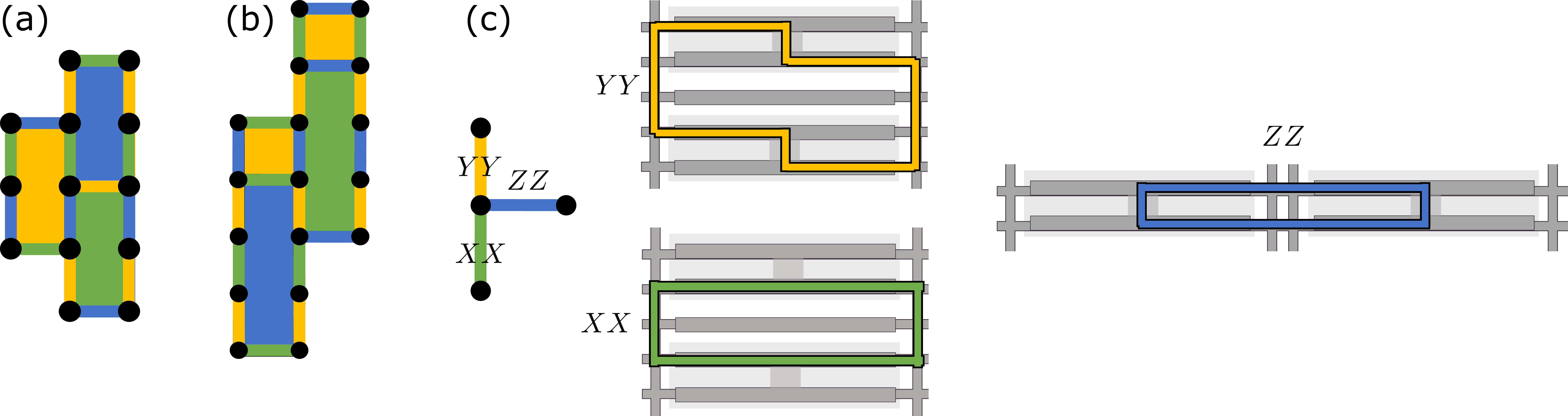}
    \caption{\label{fig:bulk_msts} 
    (a) Vertical brick wall unit of the honeycomb lattice.  (b) Vertical brick wall unit of the 4.8.8 lattice.  (c) Bulk measurement loops.
    Colors refer to the measurement time steps, rather than the Pauli operators.  
    }
\end{figure*}

Single- and multi-qubit Pauli measurements are performed by forming an interference loop that includes the corresponding set of MZMs. These interference loops must be formed such that electrons can travel along them coherently, with the loop entering and exiting the qubit(s) through the MZM pair(s) of interest and the remainder of the loop formed by semiconducting segments. The semiconductor is shown in tan in~\cref{fig:phys-layout}; not shown are tunable junctions that allow interference loops to be configured on nanosecond timescales. If the loops additionally include semiconductor quantum dots, the state of the qubits will shift the excitation spectrum of the quantum dot in a way that can be detected using standard microwave frequency techniques, thus performing a measurement of the (multi-)qubit Pauli operator~\cite{Karzig2017}.
To form long loops (in excess of the semiconductor phase coherence length), it is convenient to introduce additional floating topological wires, so-called `coherent links'~\cite{Karzig2017}, which can be seen as single topological wires between the tetron qubits in \cref{fig:phys-layout}.
While in principle measurement loops can be made arbitrarily large, in practice the measurement error increases with the size of the loop~\cite{Knapp18,Tran20}, and it is thus desirable to rely on measurements corresponding to short loops. The length of the loop for a given Pauli measurement can be optimized by choosing from the set of equivalent Majorana operators the one that leads to the shortest loop.
For this specific layout, two-qubit measurements that avoid coherent links ($ZZ$, $XX$, $XY$, $YX$, and $YY$ between vertical nearest neighbors, $ZZ$ between horizontal nearest neighbors) are expected to be of similar fidelity to single-qubit measurements with one coherent link ($X$ and $Y$), and higher than those involving additional superconducting islands (e.g. $ZX$ between vertical or horizontal neighbors).
Note that only one pair of MZMs on the qubit can be measured at a time in order to define a single path through the qubit island, and that measurement loops cannot intersect.

The honeycomb and 4.8.8 lattices map to the tetron array in \cref{fig:phys-layout} using vertical bricks such that each $2n$-gon corresponds to a rectangle of height $n$. 
With this mapping, the honeycomb and 4.8.8 Floquet codes in the bulk use a subset of the highest fidelity two-qubit measurements: $XX$ and $YY$ measurements between vertically adjacent qubit islands, and $ZZ$ between horizontally adjacent qubit islands, see~\cref{fig:bulk_msts}.
To implement measurements on neighboring pairs of vertically adjacent qubits without their corresponding loops intersecting, we use two columns of semiconductor separating each column of qubits.
We thus see that the implementation of both the honeycomb and 4.8.8 Floquet codes on this physical layout does not require any auxiliary qubits or extra time steps, a significant advantage compared to codes built out of higher-weight stabilizers~\cite{Tran20}.

\begin{figure*}[t]
    \centering
    \includegraphics[width=2\columnwidth]{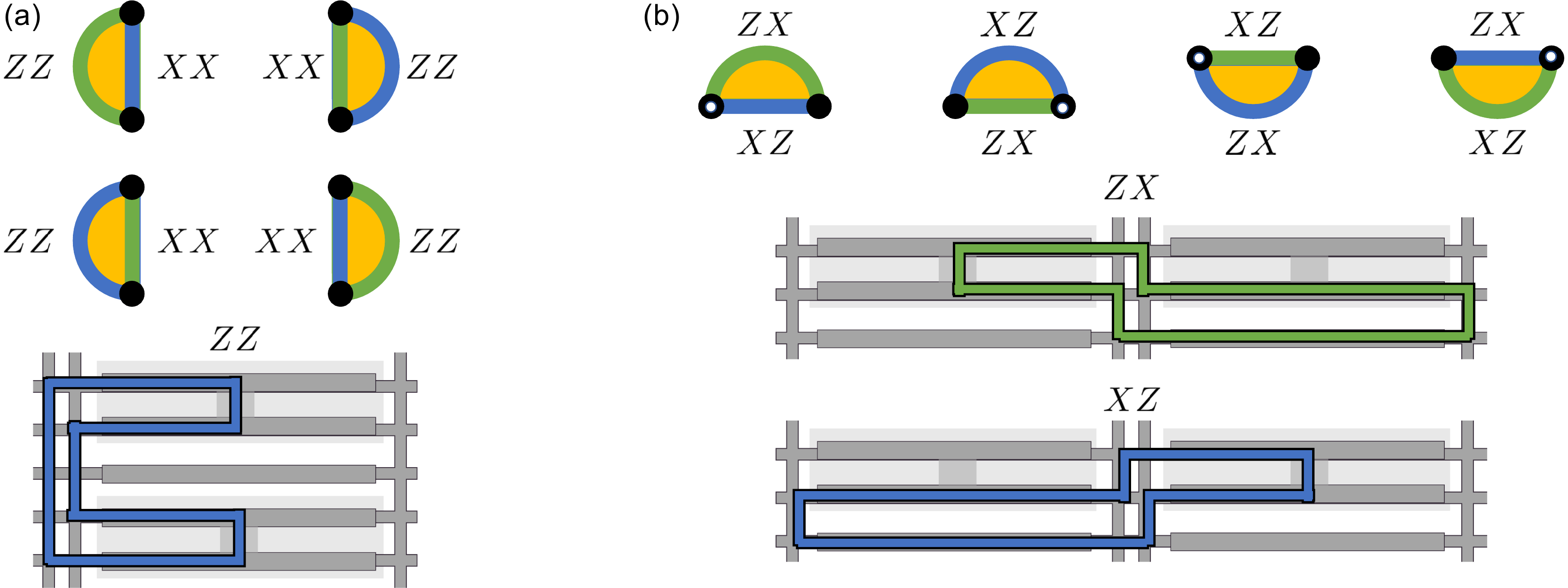}
    \caption{\label{fig:488_bdry_msts} Boundary measurements for the 4.8.8 Floquet code.  
    (a) 2-gons along vertical boundaries require $ZZ$ between vertically adjacent qubit islands, which can be implemented using both semiconducting columns to the left or right of the qubit.
    (b) 2-gons along horizontal boundaries require $ZX$ and $XZ$ between horizontally adjacent qubit islands, which can be implemented using the coherent links above or below the qubit islands.
    The white dot indicates that, for convenient physical implementation, the role of $X$ and $Z$ operators are swapped for that qubit as compared to the bulk.  For example, a horizontal edge is assigned $ZZ$ in the bulk, but becomes either $XZ$ or $ZX$ when an endpoint is marked by the white dot.  
    }
\end{figure*}

To implement the boundary conditions depicted for the code patches in~\cref{fig:honeycomb-parallelogram} and \cref{fig:488-rectangle-2x2} requires additional measurements.
In the 4.8.8 Floquet code, extending the bulk measurement pattern to the boundaries introduces $ZZ$ between vertically adjacent qubit islands, and $XX$ between horizontally adjacent qubit islands.  
While the former can be implemented without using any coherent links \cref{fig:488_bdry_msts}(a), the latter requires two coherent links. 
To avoid this, we modify the measurement pattern for one of the qubits forming 2-gons along horizontal boundaries.
One choice is to swap $X$ and $Z$ for one of the qubits, indicated by a white dot in \cref{fig:bdry_msts} (b).
This modification swaps a no-coherent link ($ZZ$) and a two-coherent link ($XX$) measurement for two one-coherent link measurements ($XZ$ and $ZX$).

\begin{figure*}[t]
    \centering
    \includegraphics[width=2\columnwidth]{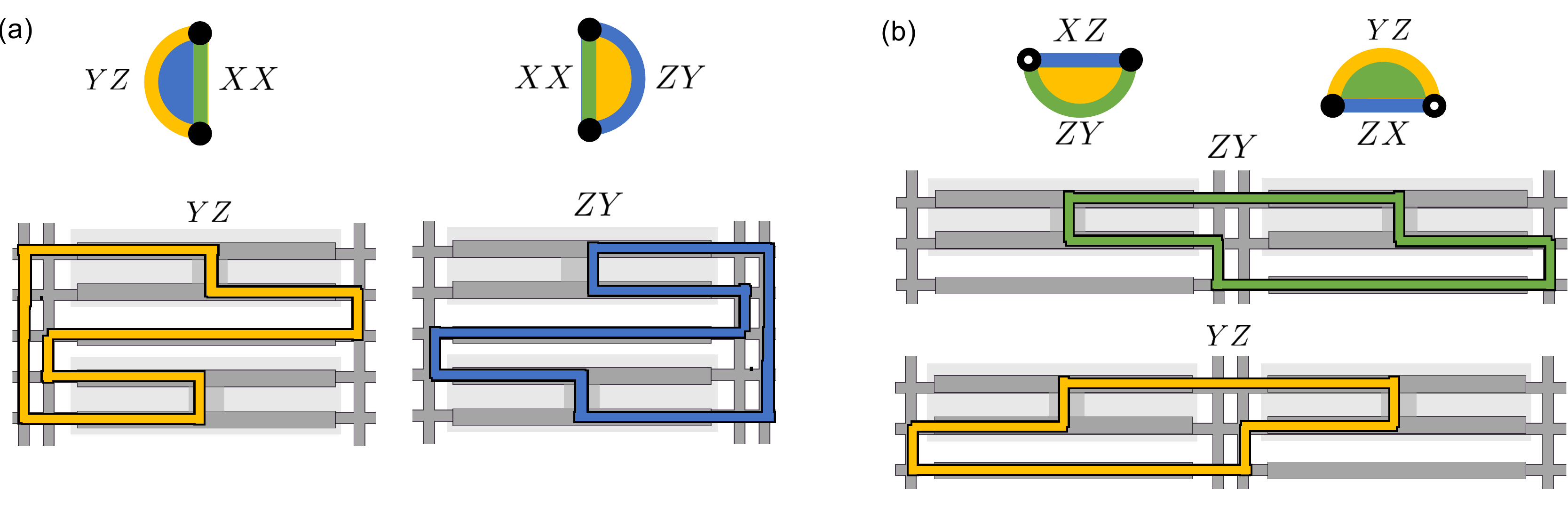}
    \caption{\label{fig:honeycomb_bdry_msts} 
    Boundary measurements for the honeycomb Floquet code.
    (a) 2-gons along vertical boundaries require $YZ$ or $ZY$ between vertically adjacent qubit islands, which can be implemented using both semiconducting columns to the left or right of the qubit and the coherent link between the qubits.
    (b) 2-gons along horizontal boundaries require $ZY$ and $XZ$ or $YZ$ and $ZX$ measurements between horizontally adjacent qubit islands, which can be implemented using the coherent links above or below the qubit islands.  
    Like~\cref{fig:488_bdry_msts}, the white dot indicates that the role of $X$ and $Z$ operators are swapped for that qubit as compared to the bulk.
    Note that $XZ$ and $ZX$ measurement loops are shown in \cref{fig:488_bdry_msts}.
    }
\end{figure*}

For the honeycomb code, extending the bulk measurement pattern to the boundaries requires measuring $ZY$ and $YZ$ between vertically adjacent qubit islands, and $YX$ and $XY$ between horizontally adjacent qubit islands.
The new vertical measurements each make use of a single coherent link, see \cref{fig:honeycomb_bdry_msts}(a).
The new pair of horizontal measurements each require two coherent links; we can once again avoid these by modifying the measurement pattern for one of the qubits forming 2-gons along horizontal boundaries (e.g. from $ZZ$ and $XY$ along the bottom boundary 2-gons to $XZ$ and $ZY$, and from $ZZ$ and $YX$ along the top boundary to $ZX$ and $YZ$ as shown in \cref{fig:honeycomb_bdry_msts}).
There are multiple choices for how to modify these boundary conditions that introduce the same number of one-coherent link measurements.

We explicitly show the full measurement circuit for the 4.8.8 Floquet code on the physical layout in \cref{app:mst-circuit}.
Note that as the 4.8.8 Floquet code only uses coherent links along certain horizontal boundary measurements, it is slightly preferable from the physical layout point of view compared to the honeycomb code.

\section{Logical performance estimates}

\begin{table}
\centering
\begin{tabular}{|l|l|l|l|}
    \hline
    Code family & Boundary & Code size & Threshold \\
    \hline
    \multirow{2}{8em}{6.6.6 Floquet (honeycomb) } & torus & $6d^2$ & 1.2 -- 1.3\% \\
    \cline{2-4} %MPS
    & parallelogram & $6d^2 + 4(d-1)$ & 1.1 -- 1.3\%  \\
    \hline
    \multirow{2}{8em}{4.8.8 Floquet (truncated square)} & torus & $4d^2$ & 1.0 -- 1.2\%  \\
    \cline{2-4} %MPS
    & rectangle & $4d^2 + 8(d-1)$& 1.0 -- 1.2\% \\
    \hline
\end{tabular}
\caption{
Threshold estimates for code implementations with two-qubit Pauli measurements.  The effective distance $d$ is for a circuit noise model with two-qubit correlated errors. For the 4.8.8 code, torus distance must be even and
rectangle distance must be odd.
Note that, for square grid of Majorana-based qubits, the code size coincides with the number of qubits required for implementation.
}\label{table:thresholds}
\end{table}

\begin{figure*}
\subfloat[honeycomb torus]{
    \includegraphics[width=.35\textwidth]{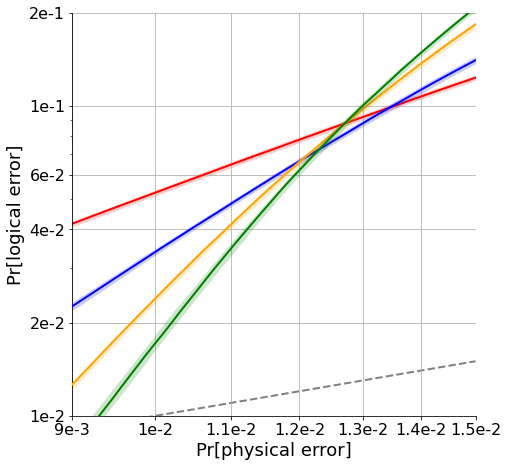}
    \label{fig:honeycomb-curves-torus-012012}
}
\subfloat[honeycomb parallelogram]{
    \includegraphics[width=.35\textwidth]{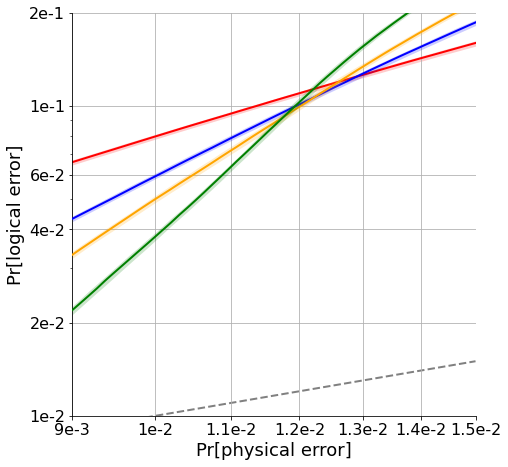}
    \label{fig:honeycomb-curves-paralellogram}
}
\includegraphics[width=.1\textwidth]{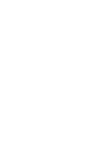}
\hfill
\subfloat[4.8.8 torus]{
    \includegraphics[width=.35\textwidth]{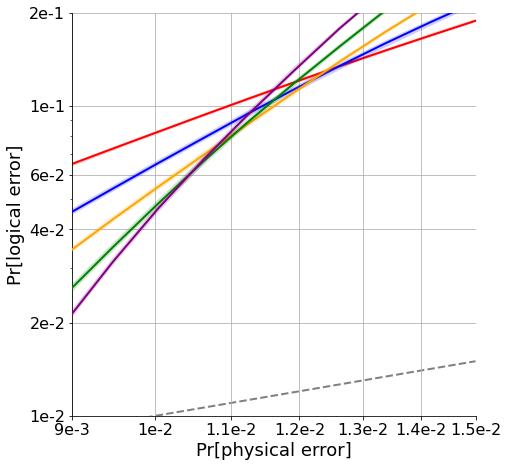}
    \label{fig:488-torus-012012}
}
\subfloat[4.8.8 rectangle]{
    \includegraphics[width=.35\textwidth]{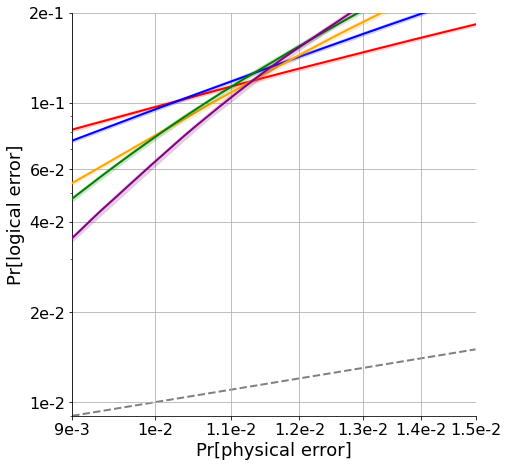}
    \label{fig:488-paralellogram}
}
\subfloat{
    \includegraphics[width=.1\textwidth]{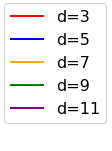}
}
\caption{
Logical error rate estimates for the honeycomb and 4.8.8 Floquet codes near threshold.  In the legend, $d$ is the effective distance of the code. The dashed grey line indicates Pr[logical error]$=$Pr[physical error].
}
\label{fig:threshold-curves} 
\end{figure*}

\begin{figure*}
\includegraphics[width=.2\textwidth]{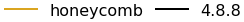}
\subfloat{
    \includegraphics[width=.31\textwidth]{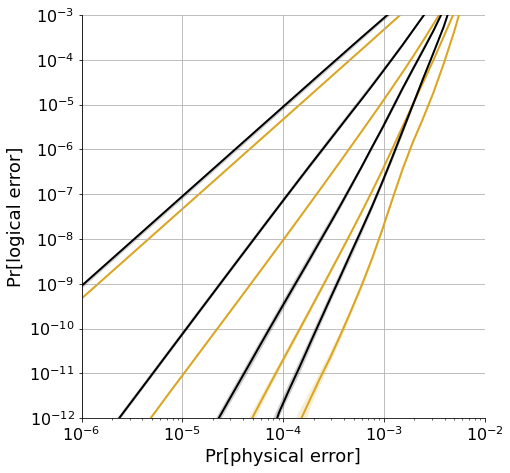}
    \label{fig:logical-curves}
}
\subfloat{
    \includegraphics[width=.31\textwidth]{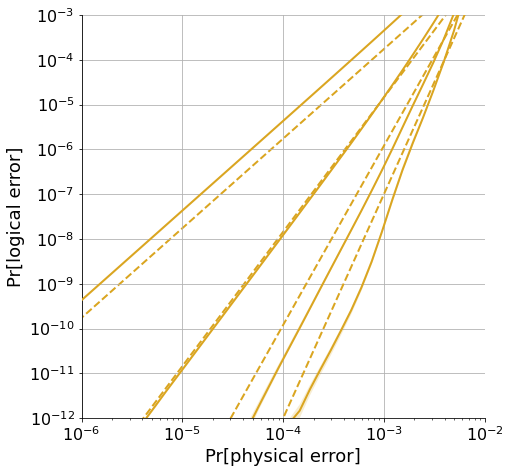}
    \label{fig:honeycomb-heurisic-curves}
}
\subfloat{
    \includegraphics[width=.31\textwidth]{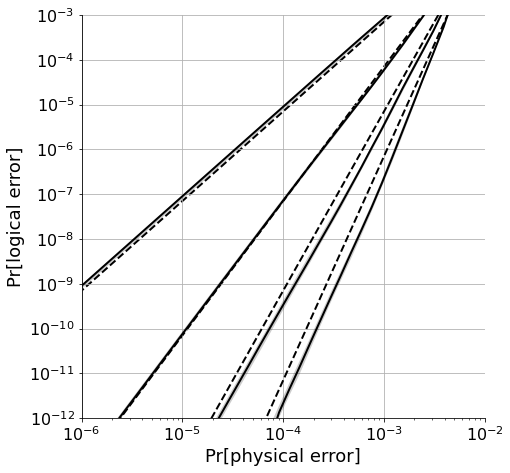}
    \label{fig:488-heuristic-curves}
}
\caption{\label{fig:low-error-estimates}
(a) Empirical logical error estimates for the honeycomb and 4.8.8 Floquet codes with planar boundaries.  Effective distances (3,5,7,9) are shown.
(b) Heuristic estimates of $0.25 (p / 0.012)^{(d+1)/2}$ as a function of effective distance $d$ for the parallelogram honeycomb code.  Empirical estimates are shown as solid lines for reference.
(c) Heuristic estimates of $0.07 (p / 0.01)^{(d+1)/2}$ for the rectangular 4.8.8 Floquet code.
}
\end{figure*}

\begin{figure*}
\hspace{.1\textwidth}
\includegraphics[width=.6\textwidth]{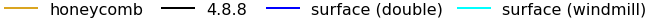}    
\hspace{.1\textwidth}
\subfloat{
    \begin{tabular}{c}
       \includegraphics[width=.25\textwidth]{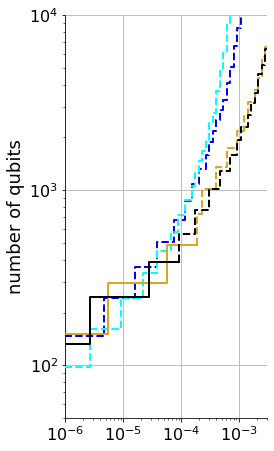}\\
       \includegraphics[width=.25\textwidth]{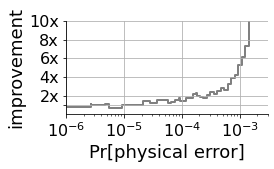}
    \end{tabular}
}
\hfill
\subfloat{
    \begin{tabular}{c}
       \includegraphics[width=.25\textwidth]{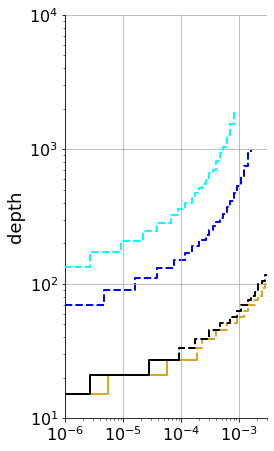}\\
       \includegraphics[width=.25\textwidth]{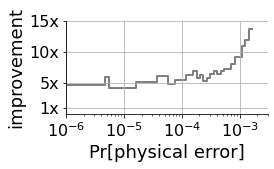}
    \end{tabular}
}
\hfill
\subfloat{
    \begin{tabular}{c}
       \includegraphics[width=.25\textwidth]{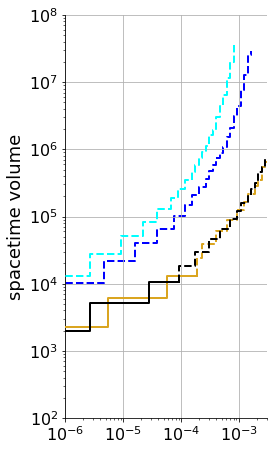}\\
       \includegraphics[width=.25\textwidth]{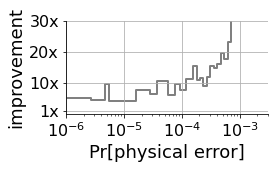}
    \end{tabular}
}
\caption{\label{fig:overhead-curves}
Overhead estimates for a target logical error rate of $10^{-12}$ with respect to (a) number of qubits (space) (b) measurement depth (time) and (c) spacetime volume (number of qubits $\times$ measurement depth).
Solid lines are based on empirical estimates from~\cref{fig:logical-curves}, dashed lines are based heuristic estimates from~\cref{fig:honeycomb-heurisic-curves,fig:488-heuristic-curves}. 
For reference, corresponding surface code estimates from~\cite{Chao2020} are shown in blue.
The gray lines along the bottom show the improvement factor of the Floquet codes relative to the surface code. They are obtained by taking the minimum of the two surface code values (double or windmill), the minimum of the two Floquet code values (honeycomb or 4.8.8) and computing the ratio surface/Floquet.
}
\end{figure*}

We now turn to numerical simulations of the logical performance of the codes described in the previous sections. All results use a gate set consisting of direct two-qubit Pauli measurements, and a two-qubit depolarizing noise model as described in~\cref{sec:computational-methods}. Our results for the thresholds are summarized in~\cref{table:thresholds}. Logical performance 
estimates for the honeycomb and 4.8.8 Floquet codes are shown in~\cref{fig:threshold-curves} and~\cref{fig:logical-curves}, where the former focuses on the threshold region for each code family while the latter shows the performance far below threshold.

\subsection{Thresholds}
Our performance estimates allow for rough comparison of thresholds between periodic and planar boundary conditions, and also between honeycomb and 4.8.8  Floquet codes. For the honeycomb code the planar threshold of $\sim1.1\%$ is somewhat smaller than the toric threshold of $\sim1.3\%$.  It is unclear if these differences are meaningful, as more extensive simulations would be needed to accurately estimate the thresholds and their separations.
For the 4.8.8 code, planar and toric thresholds are roughly comparable, both around $\sim 1.1\%$.

The more significant separation in thresholds is between measurement-based surface code implementations and the Floquet codes described here.
In Ref.~\onlinecite{Chao2020}, implementations of the surface code with two-qubit Pauli measurements were shown to have a threshold of $\sim0.2\%$ for the same noise model used here.  This large difference\footnote{Ref.~\cite{Chao2020} used union-find decoders which could also contribute to the difference in thresholds, but we do not expect the different decoders to lead to such a large separation.} is likely due to the large circuit volume for the surface code---each plaquette measurement in the surface code requires ten time steps, while the honeycomb code requires only two time steps and three measurements (on average) for each plaquette.

Finally, our estimate for the honeycomb threshold on the torus is lower than the 1.6\% -- 2.0\% reported in Ref.~\cite{Gidney2021}.  We speculate this could be due to differences in methods for computing logical error rate, details of the decoder, or code geometry.  For example, our simulations use $d/2$ rounds of noisy measurements flanked by rounds of ideal measurements (see~\cref{sec:calculating-logical-errors}), whereas Ref.~\onlinecite{Gidney2021} uses $3d/2$ rounds of noisy measurements.

\subsection{Sub-threshold performance}\label{sec:low-error-rate-performance}
While threshold estimates are of theoretical interest, resource requirements for fault-tolerant quantum computers will be governed by logical error performance for physical errors well below threshold.  Of particular interest in this regime is the relative performance between the two Floquet codes, and also as compared to the surface code.

Empirical logical error estimates for planar versions of the honeycomb and 4.8.8 Floquet codes at physical noise rates below threshold are given in~\cref{fig:logical-curves}.  We omit estimates for the torus since periodic boundary conditions are not expected to be practical at large scale.  
For effective distances beyond our empirical data we use the threshold-based heuristic method of~\cite{Fowler2012}.
For a threshold estimate $p_{th}$ and fixed constant $C$ the logical error rate for effective distance $d$ is estimated as
\begin{equation}\label{eq:logical-error-heuristic}
\Pr[\text{logical}] \approx C \left(\frac{p}{p_\text{th}}\right)^{(d+1)/2}.
\end{equation}

We choose $C$ so that~\cref{eq:logical-error-heuristic} upper-bounds our empirical estimates for $d > 3$ and $p \leq 10^{-3}$, see~\cref{fig:honeycomb-heurisic-curves} and~\cref{fig:488-heuristic-curves}.
For a fixed distance, the sub-threshold logical error rates between the honeycomb and 4.8.8 codes can differ substantially. Except for small distances, the honeycomb code offers a lower logical error rate.

\subsection{Space and time overhead}\label{sec:space-overhead}
From our logical error estimates we may estimate the space (number of qubits) and time requirements of the Floquet codes. The overhead estimates for a target logical error rate of $10^{-12}$ are given in~\cref{fig:overhead-curves}.  
For reference, we compare against corresponding resource requirements for measurement-based implementations of the surface code, as given in~\cite{Chao2020}.

Time overheads for the Floquet codes are strictly and significantly better than those of the surface code.  We estimate between a $5\times$ and $10\times$ improvement in time overhead for physical error rates between $10^{-6}$ and $10^{-3}$. The time savings is a consequence of the differences in measurement depth per logical cycle.  For the surface code, measurement depth is either $10d$ for the ``double-ancilla'' layout or $19d$ for the ``windmill'' layout.  Compare this to a corresponding measurement depth of $3d$ for the Floquet codes.  In addition, due to lower logical error rates, the Floquet codes require smaller effective distance $d$.

Space savings over the surface code can be up to $10\times$ for error rates above $0.1\%$.  However, for a fixed distance, the surface code requires fewer qubits ($2d^2$ or $3d^2$) than the Floquet codes ($4d^2 + 8(d-1)$ or $6d^2 + 4(d-1)$).
Accordingly, the savings decrease with the physical error rate.  
For physical error between $10^{-5}$ and $10^{-3}$, space savings are between $1$--$5\times$.
For physical error between $10^{-6}$ and $10^{-5}$, space requirements slightly favor surface codes.

For a model in which errors occur independently on each qubit the effective distance of the honeycomb code doubles, while the effective distance of the 4.8.8 code (and the surface code) remains unchanged, see~\cref{sec:honeycomb-codes}.  Under that model, the space overhead of the honeycomb code could be substantially better than the 4.8.8 code and the surface code.
Another option for the honeycomb code is to use ancilla qubits and multiple measurements to implement each check operator.  This mitigates two-qubit errors and doubles the effective distance in the presence of two-qubit correlated noise. But it also requires more measurements, increasing both the time per logical operation and circuit volume.  We do not consider that case here.

\section{Computational methods}\label{sec:computational-methods}

\subsection{Noise model}\label{sec:noise-model}
We use the noise model defined in~\cite{Chao2020}, and coined ``EM3'' in~\cite{Gidney2021}.  In this model, each two-qubit measurement fails independently with probability $p$.  When it fails, it acts as the ideal measurement followed by an error drawn uniformly from the set of non-trivial errors $\{P_1 \otimes P_2 \times F\} - \{I\otimes I \times 0\}$, where $P_1,P_2 \in \{I,X,Y,Z\}$ are Pauli errors acting on the support of the measurement and $F\in\{0,1\}$ is a bit flip of the measurement outcome.

This noise model is chosen for simplicity and for comparison against existing numerical simulations. We do not expect it to accurately represent noise characteristics of a physical device.

\subsection{Calculating logical errors}\label{sec:calculating-logical-errors}
To estimate the probability of a logical error, we sample errors from $d/2$ (where $d$ is the effective distance of the code) rounds of noisy syndrome measurements flanked both before and after by two rounds of ideal syndrome measurements.  For the torus, we alter the definition of a round to be two sequential period-three sequences.  This method is the direct analog of that used in~\cite{Chao2020}, but accounts for the fact that each of the plaquettes are measured twice per length-six syndrome round.

The effect of a set of errors is calculated by propagating the errors through the measurement circuit.  This yields a residual Pauli error on the qubits, and a set of flipped measurement outcomes.
The measurement flips are used as input to the decoder (see~\cref{sec:decoding}) in order to obtain a recovery Pauli operator.  The recovery operator is multiplied into the residual error of the effect to obtain a net effect.  The net effect is considered to be a logical error if exactly one of the following is true:
\begin{enumerate}
    \item the net residual Pauli error anti-commutes with one or both of the (instantaneous) logical operators or,
    \item the intersection of the bit flip errors with the set of logical phase conditions has odd size.
\end{enumerate}

The logical operators and their corresponding logical phase conditions can be computed by using stabilizer simulation of the check measurements to compute the instantaneous stabilizer group.  Pauli support (without dependent phases) can be computed from the normalizer.  Each element logical operator is then added to the instantaneous stabilizer group and used as input to the check measurement simulation in order to determine the set of measurement outcomes in the logical phase conditions.

Toric versions of the honeycomb and 4.8.8 Floquet codes each have two logical qubits and therefore four logical operators.  For fair comparison with the planar versions, which have only one logical qubit, we arbitrarily choose one of the two qubits and ignore the other.

\subsection{Decoding}\label{sec:decoding}
We use a minimum-weight perfect-matching decoder with weighted edges~\cite{dennis2002}. Edge weights are assigned based on a fixed physical error rate $p$. Therefore the performance of the decoder depends on the choice of value for $p$. For threshold estimates (\cref{fig:threshold-curves}), we assign edge weights using $p=2\%$, and for low-error estimates (\cref{fig:logical-curves}) we assign edge weights using $p=0.1\%$.

The input to the decoder is a sequence of bits that indicate changes to the eigenvalues of the stabilizers. For stabilizers (faces) in the bulk of the lattice each bit is obtained by taking the parity of adjacent check measurements over a length-six time window, i.e., two consecutive rounds of checks.  For the honeycomb code, this is the parity of 12 measurement outcomes, and for the 4.8.8 code it is the parity of either 8 or 16 measurement outcomes.
At boundaries, there are different types of input bits to the decoder, corresponding to transient stabilizers as explained in~\cref{sec:honeycomb-codes,sec:488-codes}.

The time steps for which the value of a given face can be inferred is determined by the face color.  A face of a given color is inferred only when all of its adjacent edges were measured in the previous two time steps.  For example, in the repeated check sequence (yellow, blue, green, yellow, green, blue), the corresponding faces that can be inferred are (green, green, yellow, blue, blue, yellow). Faces around the boundary follow a similar inference rule, but where the inference may span more than two time steps, see also~\cref{sec:honeycomb-codes} and~\cref{sec:488-codes}.

\subsection{Weighting set sampling}

To estimate logical error rate,
we use an importance sampling technique akin to malignant set sampling~\cite{Aliferis2007b}. 
For simplicity of explanation, suppose each location in spacetime
undergoes an error channel $\rho \mapsto (1-p)\rho + p E_\ell \rho E_\ell$
where $E_\ell$ is a Pauli operator that depends on the location~$\ell$.
Then, the overall logical error rate is given by
\begin{equation}\label{eq:uniform-logical-error-polynomial-estimate}
    \Pr[\text{logical}] = \sum_{w \ge 0} f_w \binom{n}{w} p^w (1-p)^{n-w}
\end{equation}
where $w$ is the number of locations where $E_\ell$ has occurred, which we call ``degree,''
$n$ is the total number of locations
and $f_w$ is the fraction of those that result in a logical error among all error configurations of degree~$w$.
If we uniformly sample over all possible error configurations of total degree~$w$
with sample count~$s_w$,
then we can estimate the fraction~$f_w$ by~$c_w/s_w$
where $c_w$ is the count of post-decoder states that is different from the error-free state.

Estimating the logical error in this way 
is most beneficial in the low error regime where the expected logical failure rate is very small.
The usual Monte Carlo sampling approach would require a very large sample number,
whereas the method described here does not depend on the absolute logical failure rate.

In~\cref{eq:uniform-logical-error-polynomial-estimate} the probability $p^w(1-p)^{n-w}$ of a degree-$w$ error configuration depends only on $w$.
With more general error channels that are possibly nonuniform across spacetime, the probability of a degree-$w$ error configuration also depends on the types of errors.
For example, if an error channel is 
$\rho \mapsto (1-3p)\rho + p E_1 \rho E_1 + 2p E_2 \rho E_2$,
which is the same for all locations,
then we estimate each of fractions~$f_w(j)$ where $j=0,1,2,\ldots,w$
such that
\begin{align}
    \Pr[\text{logical}] = \sum_{w \ge 0} \sum_{j=0}^w f_w(j) n_w(j) p^j (2p)^{w-j} (1-3p)^{n-w}
\end{align}
where $n_w(j)$ is the total number of error configurations of degree~$w$
with exactly $j$ occurrences of~$E_1$.
The number $n_w(j)$ can be computed by expanding a polynomial $(1+x_1+x_2)^n$
and reading off the coefficient of $x_1^j x_2^{w-j}$.
Note that one can expedite computing $n_w(j)$
by truncating intermediate polynomials and keeping only the terms up to degree~$w$.

\subsection{Bounds and statistical uncertainty}

To compute the statistical uncertainty of~\cref{eq:uniform-logical-error-polynomial-estimate}, we use a Bayesian approach with conjugate priors.  We treat each fraction $f_w = c_w/s_w \in [0, 1]$ as a random variable, where $c_w$ follows a $\mathrm{Binomial}(n=s_w,p=f_w)$ distribution, and we aim to estimate the parameter $f_w$ for each $w$. 

The total number of samples $s_w$ at each weight $w$ is chosen via a heuristic that guarantees sufficiently small variance in the estimated logical error rate at some chosen physical error rates. Taking the estimates of $f_w$ to be independent for different $w$, the variance in the estimate of the logical error rate is given by
\begin{align}
    \sigma^2_{\text{logical}} = \sum_w \sigma^2_w \left[{n\choose w}p^w(1-p)^{n-w}\right]^2.
\end{align}
Fixing $p$, we choose to equalize the contributions from each different $w$, which lead to having $s_w$  be distributed across $w$ in a manner that is inversely proportional to $\left[{n\choose w}p^w(1-p)^{n-w}\right]^2$, with a total number of samples across all $w$ set so that $\sigma_\text{logical}$ is a small fraction of $\Pr[\text{logical}]$. Since $\Pr[\text{logical}]$ is not known a priori, preliminary simulations are used to make a coarse estimate of the threshold $p_{\text{th}}$.  We then use the heuristic~\cref{eq:logical-error-heuristic}
and adjust the $s_w$. For several choices of $p$ and anticipated logical error rates, we can combine the different $s_w$ for a fixed $w$ by taking the maximum $s_w$ across different $p$ of interest.

Once the $s_w$ are fixed and the data is collected, we have all the parameters needed to compute the uncertainty in the logical error rate.  For $w < w_{\min} = (d+1)/2$ we set $f_w = 0$, since errors of those weights are always correctable. 
For $w \geq w_{\min}$ we take the prior distribution for each $f_w$ to be independent and given by $\mathrm{Beta}(\alpha=1,\beta=1)$.  
The beta distribution is conjugate to the binomial distribution, in the sense that, given a beta-distributed prior over the $f_w$, an exact Bayesian update with a binomial-distributed observation leads to a beta-distributed posterior over the same parameter~\cite{Murphy2012}. More concretely, after sampling we have a set of $(c_w, s_w)$ pairs from which we obtain a posterior distribution $\mathrm{Beta}(1 + s_w - c_w, 1 + c_w)$ for $f_w$.  

Following this procedure for each $w$ of interest, we may sample from the posteriors to generate samples for the polynomial in~\cref{eq:uniform-logical-error-polynomial-estimate} based on the posterior distribution for each $f_w$, and from these samples we may generate uncertainty regions based on the empirical distribution for the value of the polynomial at each physical error rate.  We generate 100 samples of the polynomial and report the empirical quartiles as a function physical error rate.

For sufficiently large $w$, sampling becomes impractical and unnecessary.  Instead, we truncate and sample only up to some fixed weight $W$.
The probability mass of the un-sampled weights can be upper bounded by $1 - \sum_{w\ge w_{\min}}^W \binom{n}{w} p^w (1-p)^{n-w}$.  We then add this bound to our estimate to obtain
\begin{equation}\label{eq:bounded-uniform-logical-error-polynomial-estimate}
    \Pr[\text{logical}] \approx 1 - \sum_{w\ge w_{\min}}^W \left(1 - f_w\right) \binom{n}{w} p^w (1-p)^{n-w}.
\end{equation}
The regions reported in~\cref{fig:threshold-curves} and~\cref{fig:logical-curves} correspond to the upper quartile from samples of~\cref{eq:uniform-logical-error-polynomial-estimate} and the lower quartile from samples of~\cref{eq:bounded-uniform-logical-error-polynomial-estimate}.  
At the scales used in~\cref{fig:threshold-curves,fig:logical-curves} the curves for \cref{eq:uniform-logical-error-polynomial-estimate,eq:bounded-uniform-logical-error-polynomial-estimate} are indistinguishable and depicted by the solid lines in the plots.

\section{Conclusion}

We introduce a new Floquet code using a 4.8.8 lattice, and analyze the performance of the planar variant of this code and the honeycomb code~\cite{Hastings2021} in measurement-based Majorana quantum computing architectures. 
We find the corresponding error threshold to be greater than 1\%.  Sub-threshold logical error rates and space and time requirements are also lower.  Compared to the surface code implementation in Ref.~\onlinecite{Chao2020}, the planar Floquet codes described here have a threshold that is an order of magnitude higher, and time and space overheads that are up to an order of magnitude smaller for many noise regimes.
This makes the honeycomb and 4.8.8 Floquet codes leading candidates for large-scale fault-tolerant quantum computation on measurement-based systems.

The honeycomb and 4.8.8 Floquet codes have a zero-overhead implementation on the lattice of Majorana-based qubits in~\cref{fig:phys-layout}, that only uses short-distance measurement loops in the bulk, supplemented by intermediate-distance (requiring one coherent link) measurement loops along the boundaries.
In contrast, implementing the surface code on measurement-based topological qubits requires introducing auxiliary qubits and extra time steps to build up the weight-four stabilizer measurements~\cite{Chao2020}, while also making use of coherent links for all two qubit measurements involving data qubits~\cite{Tran20}.
As the physical error rate is expected to increase with the measurement loop distance, we would expect that the honeycomb and 4.8.8 Floquet codes to have lower physical error rates compared to the surface code when implemented on qubits of the same quality.

There are several variations of the physical layout of the Floquet codes.
First, both the honeycomb and 4.8.8 patches have a natural extension to hexons, modifying the physical layout by including an additional topological wire for each qubit.  
In the hexon implementation, it may be possible to optimize the boundary measurements to avoid using any coherent links. 
In this case, coherent links could be removed throughout the lattice (making all bulk measurement loop distances equivalent to those in the tetron layout with reduced loop distance along the boundaries), at the cost of removing the ability to fully characterize single qubits.  
Second, the two semiconducting columns separating columns of tetrons in~\cref{fig:phys-layout} could be replaced by a single semiconducting column, at the cost of either doubling the number of time steps (to avoid simultaneously measuring adjacent vertical edges) or using coherent links throughout the bulk (by mapping the honeycomb and 4.8.8 lattices to horizontal brick wall versions).

Finally, while this work was being prepared we became aware of closely related but independent work by Gidney and Newman~\cite{Gidney2022}.  They propose different boundary conditions for the honeycomb code in which some of the checks at the boundary are reduced from two-qubit measurements to single qubits.  Like~\cite{Gidney2021}, their threshold estimates are somewhat larger than ours, possibly due to differences in decoders and possibly due to different methods for estimating logical error rate.  They conclude that the surface code still outperforms the honeycomb code for superconducting qubit and similar architectures.  However, for direct measurement architectures like those with Majorana qubits, their qualitative conclusions match ours.

\section*{Acknowledgements}
We thank Craig Gidney and Michael Newman for coordinating to make their independent results available to the public simultaneously with ours~\cite{Gidney2022}.

\appendix

\section{Physical layout measurement circuits}
\label{app:mst-circuit}

\begin{figure*}
    \centering
    \begin{tabular}{cc}
    Step 1& Step 2\\
     \includegraphics[width=\columnwidth]{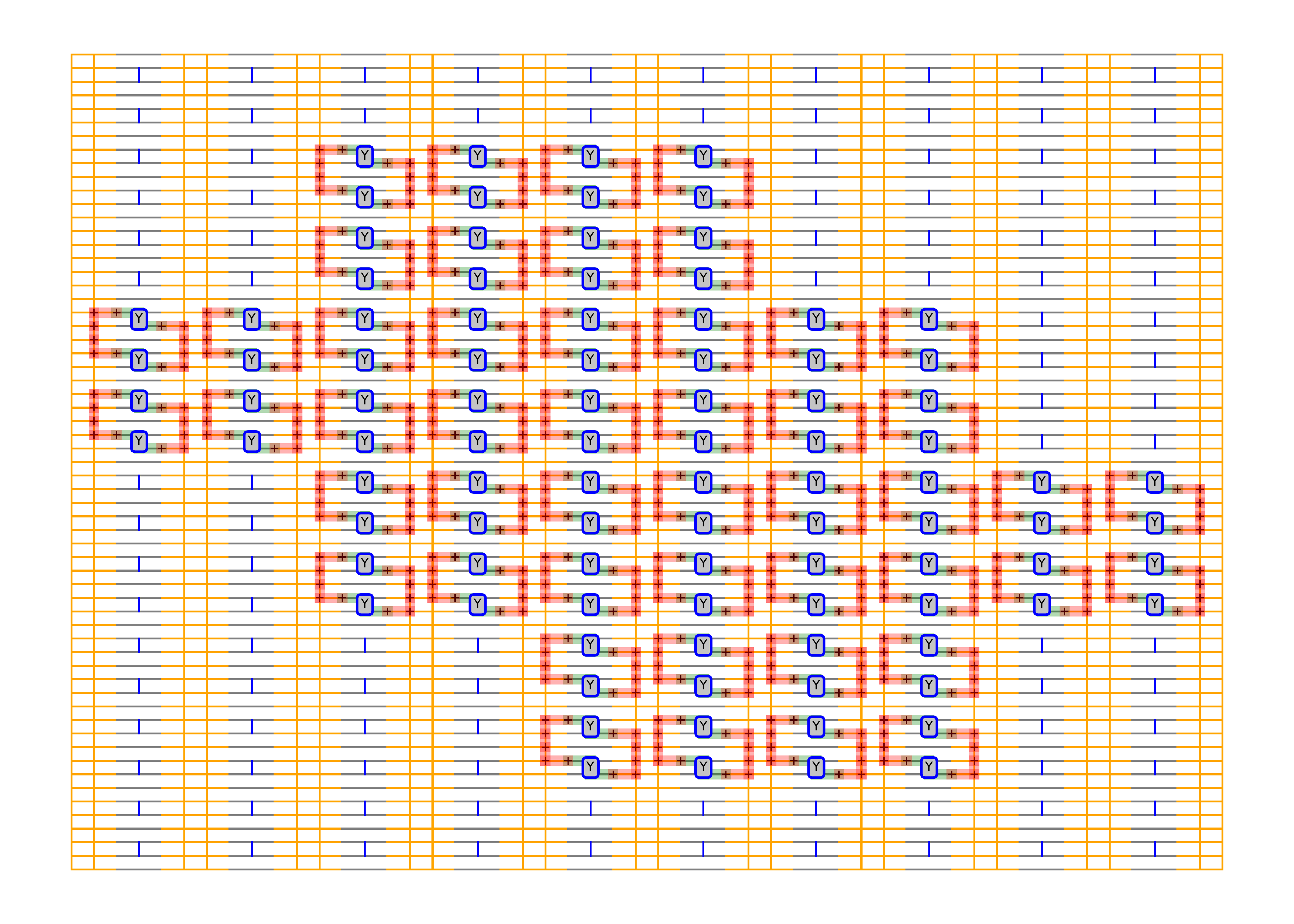}
    &\includegraphics[width=\columnwidth]{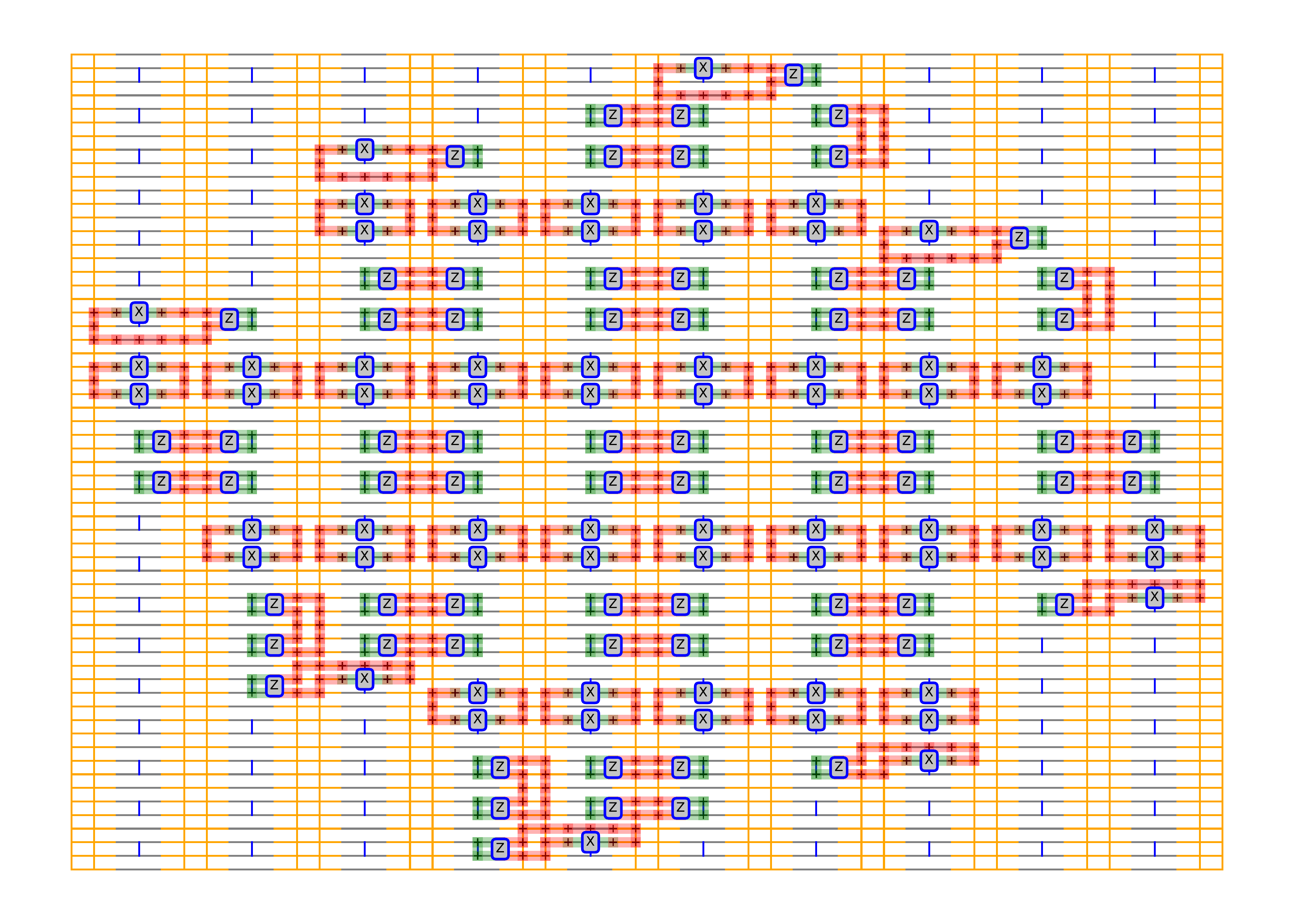} \\
    Step 3& Step 4\\
     \includegraphics[width=\columnwidth]{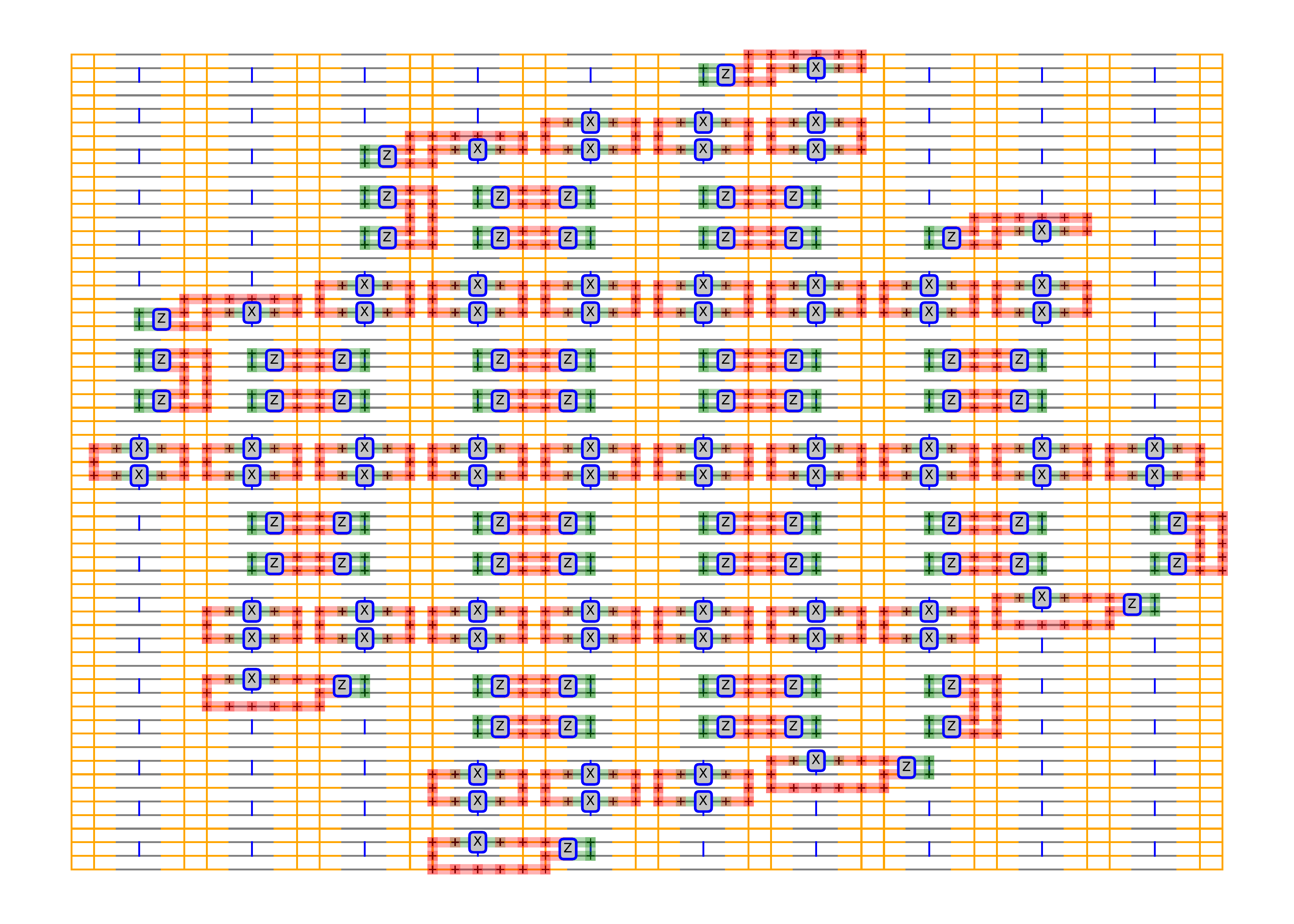}
    &\includegraphics[width=\columnwidth]{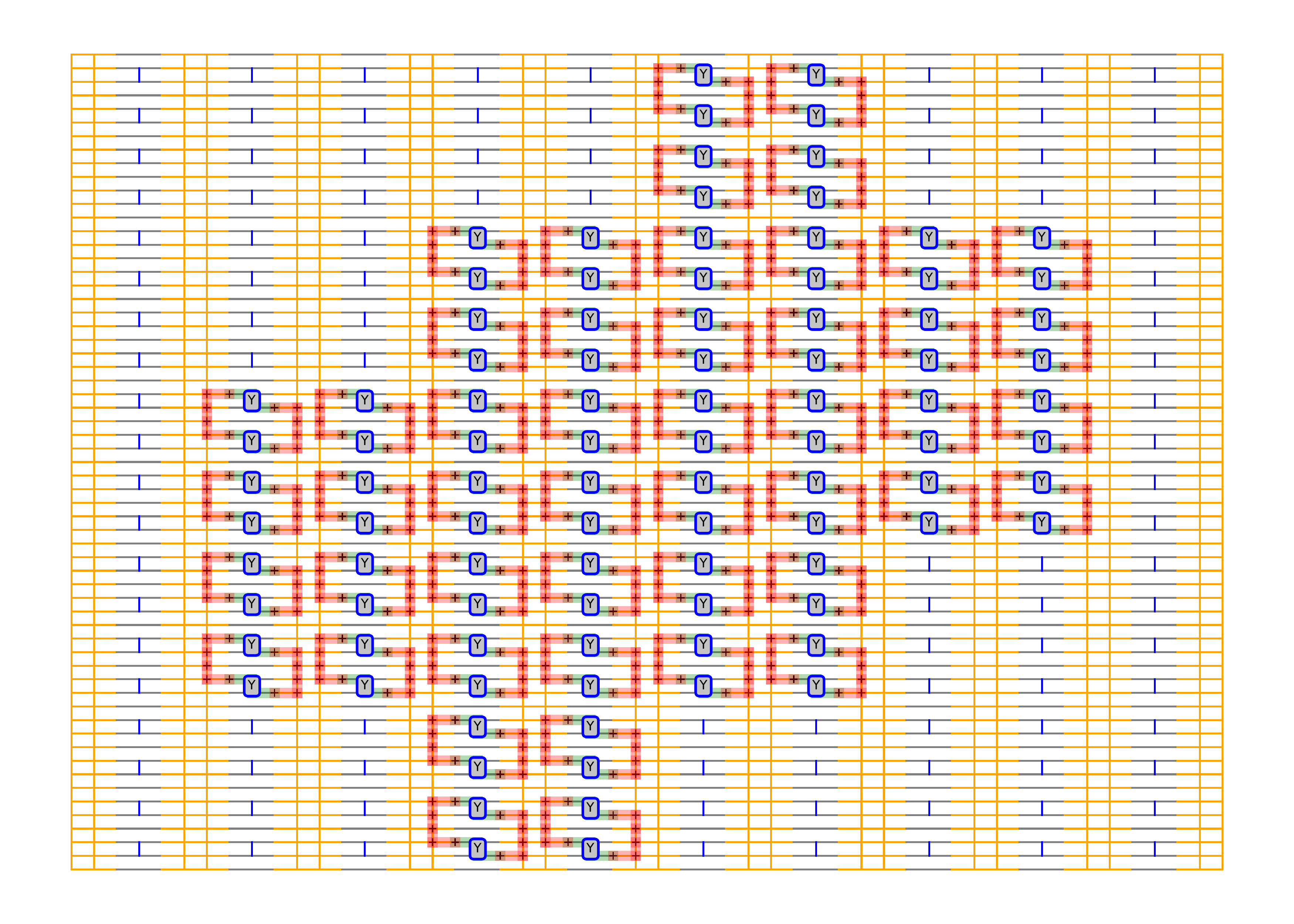} \\
    Step 5& Step 6\\
     \includegraphics[width=\columnwidth]{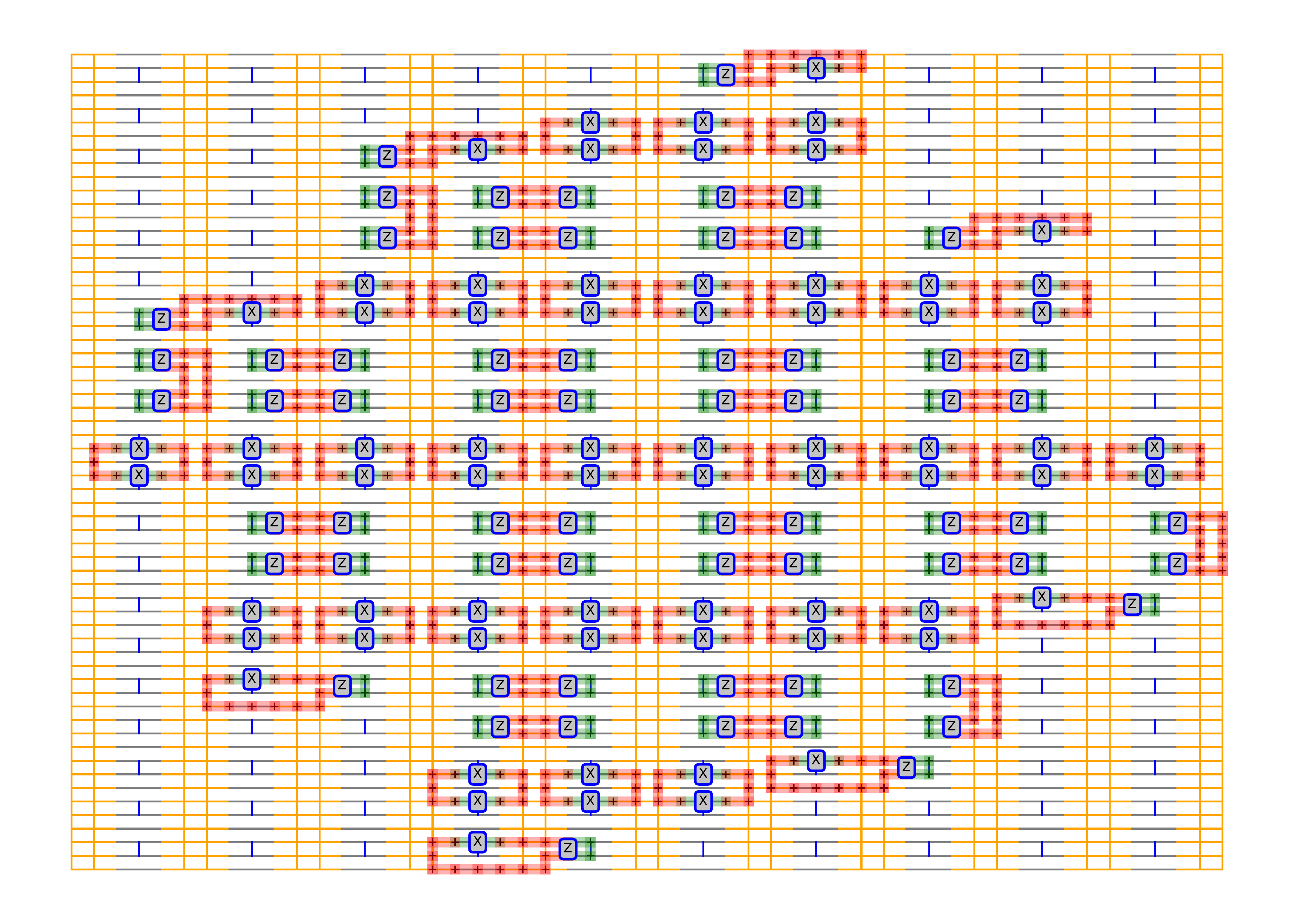}
    &\includegraphics[width=\columnwidth]{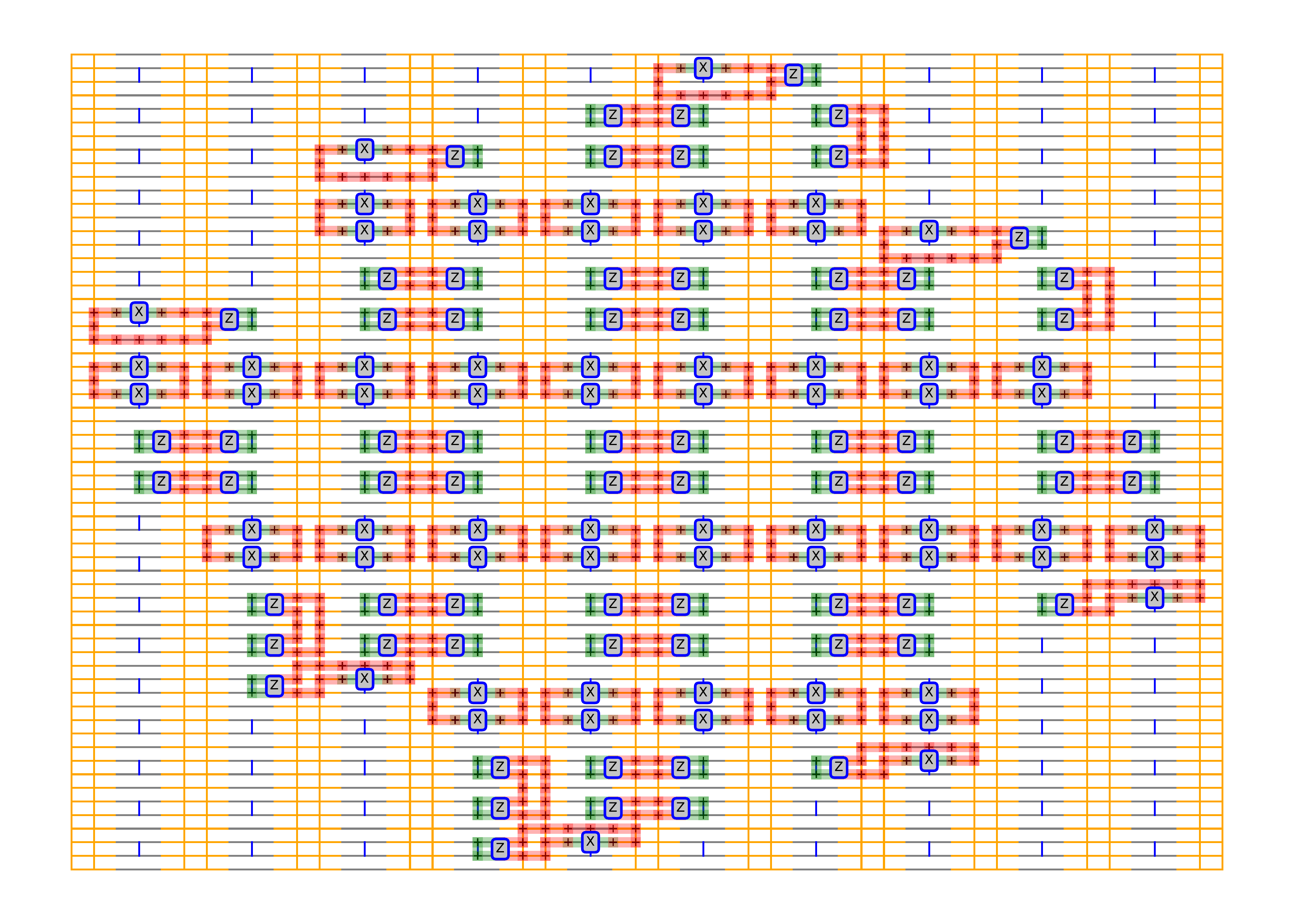}
    \end{tabular}
    \caption{Six-step measurement sequence for the planar 4.8.8 code patch shown in~\cref{fig:488-rectangle-2x2}, as implemented in an array of tetron qubits.}
    \label{fig:488-circuit}
\end{figure*}

In \cref{fig:488-circuit}, we show the six-step measurement circuit needed to implement the planar 4.8.8 code patch shown in~\cref{fig:488-rectangle-2x2} in an array of tetron qubits, as described in~\cref{sec:physical-layout}.

\bibliography{references}

\end{document}